\documentclass[preprint,showpacs,showkeys,aps,prd,amsfonts,eqsecnum,nofootinbib]
{revtex4}
\usepackage{graphicx,epsfig}
\usepackage[dvips]{color}
\newcommand{\no}{\noindent}
\newcommand{\nn}{\nonumber}

\def\gsim{\lower0.5ex\hbox{$\:\buildrel >\over\sim\:$}}
\def\lsim{\lower0.5ex\hbox{$\:\buildrel <\over\sim\:$}}
\begin{document}
\preprint{CUMQ/HEP 138}
%
%
\title{\Large RARE DECAY OF THE TOP $t \rightarrow cgg$ IN THE STANDARD MODEL}
\author{Gad Eilam}\email[]{eilam@physics.technion.ac.il}
\affiliation{Department of Physics,
Technion-Israel Institute of Technology, 32000 Haifa, ISRAEL}
\author{Mariana Frank}\email[]{mfrank@vax2.concordia.ca}
\author{Ismail Turan}\email[]{ituran@physics.concordia.ca}
\affiliation{Department of Physics, Concordia University, 7141
Sherbrooke Street West, Montreal, Quebec, CANADA H4B 1R6}
\date{\today}

\begin{abstract}
We calculate the one-loop flavor changing neutral current
top quark decay $t \to cgg$ in the Standard Model.  We demonstrate that
the rate for $t \to cgg$ exceeds the rate for a single gluon emission
$t \to cg$ by about two
orders of magnitude, while the rate for $t \to cq{\bar
q},~q=u$ is slightly smaller than for $t \to cg$.
\pacs{12.15.-y, 12.15.Ji, 14.65.Ha}
\keywords{Rare Top Decays, Standard Model, Flavor Changing Neutral
Currents}
\end{abstract}
\maketitle
\section{Introduction}\label{sec:intro}
\no
Flavor Changing Neutral Currents (FCNC)
in general, and of the top quark in
particular, play an important role as a testing ground for the
Standard Model (SM) and for New Physics (NP).
Because of its large mass, the top
quark can decay into all other quarks, accompanied
by gauge or Higgs bosons,
as well as into new particles
predicted by NP models. The interest in FCNC for
top quark physics stems from the facts that:
\begin{enumerate}
\item The scale of NP is closer to the top
quark mass more than to any other quark.
\item In the SM,
the FCNC two-body processes $t \rightarrow c g, \gamma, Z,H$
are absent at tree- level and are highly suppressed by the GIM mechanism
at one loop. Their branching ratios predicted in the SM are of the order of
$10^{-11}$ to $10^{-14}$ \cite{Eilam:1990zc,Mele:1998ag},
far away from present and even future
reaches of either the Large Hadron Collider (LHC) \cite{Beneke:2000hk,
ATLAS} or the International Linear Collider (ILC) \cite{Cobal:2004zt}.
There are many models of NP in which the branching ratios for
the above 2-body FCNC decays are much larger than those obtained in the SM
(see e.g.  \cite{Frank:2005vd,Arhrib:2005nx} and references therein).
\end{enumerate}

In addition to the two-body rare decays of top quark, some of its
rare three-body decays e.g., $t\to cWW, cZZ, bWZ$ have been
considered in the literature within the SM
\cite{Jenkins:1996zd, Altarelli:2000nt, Bar-Shalom:2005cf}
and for NP \cite{Bar-Shalom:1997sj, Bar-Shalom:2005cf}.
These three body decays are suppressed with respect
to two-body decays in the SM but some of them get
comparably large within models of NP, such as
two-Higgs-Doublet \cite{Bar-Shalom:1997sj}, especially after
including finite-width effects \cite{Bar-Shalom:2005cf}.

In this paper we will analyze another three-body rare decay, namely
$t\to cgg$ within the SM framework and compare it to
both $t\to cg$ and $t\to cq\bar{q},~q=u$. The main motivation
for such a calculation comes from the phenomenon in
which higher order dominates over a lower order rate, as observed in
the $c$, $b$ and in other systems.

For the case of charm decays,
the short distance contribution to $c\to u\gamma$, exhibits a
huge enhancement over the lowest order penguin diagrams
\cite{Greub:1996wn}. One can argue then that even higher
order short distance are not that important.
Unfortunately for radiative $D$ decays, even this enhancement
is overshadowed by much larger long distance terms.

For $b$-quark decays, a study of the next-to-leading logarithmic (NLL)
QCD corrections \cite{Greub:2000sy},
yielded ${\rm Br}^{\rm NLL}(b\to sg)\approx 5.0\times 10^{-3}$, whereas the
leading-logarithmic (LL) result
${\rm Br}^{\rm LL}(b\to sg)\approx 2.2\times 10^{-3}$, in spite of an
$\alpha_s/\pi$
suppression. This large correction is dynamical in nature, since it is
due to a large ratio of Wilson coefficients evaluated at $m_b$,
$C_2/C_8\approx 7$ (for earlier references see
\cite{{Hou:1988wt},{Hou:1990js},{Simma:1990nr},{Liu:1989pc}}),
Higher order dominance may also become substantial
in the decays of a sequential fourth generation quark $b^\prime$, if
it exists \cite{{Eilam:1989zm},{Hou:1990js},{Arhrib:2000ct}}.

More recently  \cite{Cordero-Cid:2004hk}, the one-loop, three-body,
rare top quark decay $t\to u_1 {\bar u}_2 u_2$, where $u_i=u,c$
was calculated in the SM and found to dominate over the
one-loop, two-body, rare $t\to u_1 g$  decay, by about one order of magnitude,
although the latter is of lower order in  $\alpha_s$.
Later on, we will comment about their result.

The purpose of this article is to evaluate and discuss the
higher order dominance issue in rare top
decays among $t\to cg , t\to cq\bar{q}$, and $t\to
cgg$. The present calculations are within the SM and while, as discussed
above, $t\to cg$ and $t\to c q\bar{q},~q=u$
were calculated before, this is the first
calculation of $t\to cgg$ in the SM.

The remainder of
the paper is organized as follows: In Section II we present the calculation
of $t\to cgg$, in Section III the decay $t\to c q\bar{q},~q=u$ is evaluated,
and in Section IV we conclude. The Appendix includes the one-loop functions
which appear in Section II.
\section{Calculation of $t \to cgg$}

The one-loop $t-c-g^*(k)$ and, in general, the $q_1-q_2-g^*(k)$ vertex function
can be expressed, using Lorentz and gauge invariance, as 
\cite{Deshpande:1991pn}
\begin{equation}
\Gamma_\mu = F_1(k^2)\left(k^2 \gamma_\mu -k_\mu k\!\!\!/\right)P_L -i F_2(k^2)
  m_t\, \sigma_{\mu\nu}\,k^\nu\, P_R,
\end{equation}
where $P_{L,R}\equiv (1\mp\gamma_5)/2$  and $m_c=0$ are assumed. The functions
$F_1$ and $F_2$ are called {\it charge-radius} (or {\it monopolar})
and {\it dipole moment} (or {\it dipolar}) form factor, respectively.
Note that this is not the most general
vertex function. There are two more form factors, namely $F_{1R}$ 
(right-handed monopolar)
  and $F_{2L}$ (left-handed dipolar), which are both proportional to 
$m_c/m_t$ so that
for the sake of simplicity we omit them here (see 
\cite{Deshpande:1991pn} for the details).
In our numerical analyses, however, all contributions are retained.

While $F_1(k^2)$ contribution to $q_1\to q_2g,\,\,q_1=t,\,q_2=c$ (i.e. for a
real gluon) vanishes,
both $F_1(k^2)$ and $F_2(k^2)$  give non-zero contribution to $q_1\to
q_2 q\bar{q}$
and to $q_1\to q_2 gg$. Of course, the vertex functions $\Gamma_{\mu\nu}$
for the three-body processes are more complicated than just $\Gamma_\mu$
above, but if $F_2< F_1$, there is a chance that the three-body modes
will be of the same order, or even larger than $t\to cg$.


   \begin{figure}[h]
\vspace*{-3.6in}
      \centerline{ \epsfxsize 9in {\epsfbox{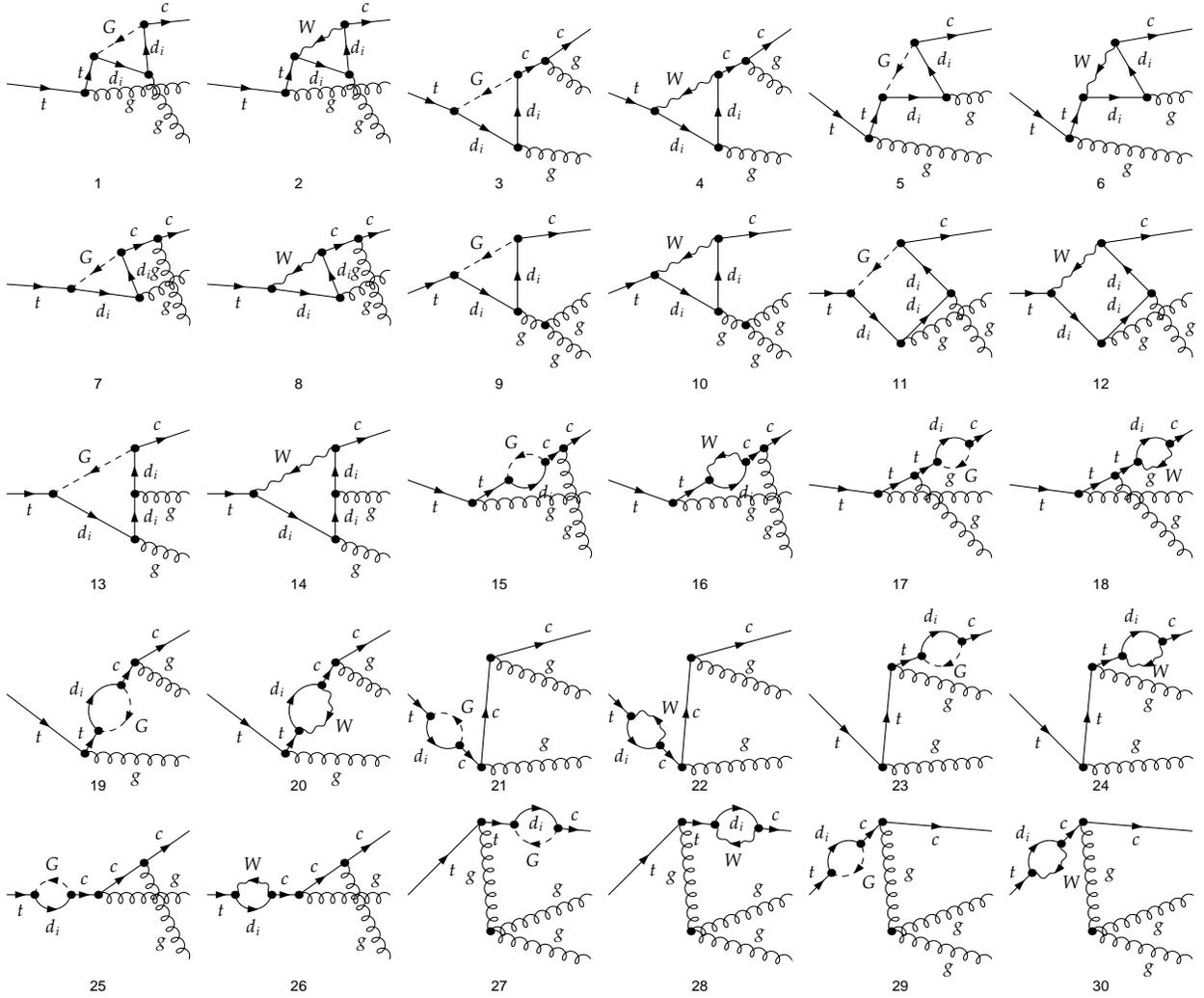}}}
\vspace*{-3.7in}
\caption
   {The one-loop contributions to $t \to cgg$ in the SM in the
't Hooft -
Feynman gauge. The ghost contributions are depicted in Fig.
\ref{Fig:tcggGhost}.
    The first 1-10 diagrams are the vertex diagrams, the diagrams 11-14 are the
one-particle irreducible (1PI or box) diagrams, and the rest 15-30  are the
    $t-c$ self energy diagrams. The crossed diagrams are also shown explicitly.}
    \label{Fig:tcgg}
\end{figure}

\begin{figure}[h]
\vspace*{-5.8in}
      \centerline{\epsfxsize 9in {\epsfbox{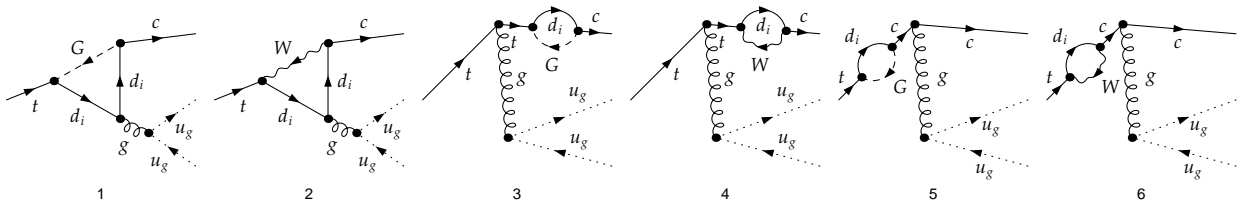}}}
\vspace*{-6in}
\caption{The one-loop ghost
contributions to $t \to cgg$ in
the SM in the 't Hooft - Feynman
gauge. There is a ghost
contribution for each diagram of
Fig.~\ref{Fig:tcgg} with triple gluon vertex.} \label{Fig:tcggGhost}
\end{figure}

In the SM, the decay $t\to cgg$ occurs at one-loop level and the
Feynman diagrams contributing to the
decay are depicted in Fig.~\ref{Fig:tcgg} in the t' Hooft - Feynman
gauge ($\xi = 1$).\footnote{Note that there will be no cross term for
diagrams with
     triple gluon vertex
     since it is already counted in the vertex factor.}
The $G$ field in the diagrams is the unphysical part of the Higgs field.
The polarization sum of the
gluons is naively
\begin{equation}
\sum_{\lambda}\epsilon_{\mu}^*(k,\lambda)\epsilon_{\nu}(k,\lambda) =
-g_{\mu\nu}.
\end{equation}
However, it is well known \cite{cutlers,field}
that in cases where there are two or more external
gluons (either for tree, like $g q\to g q$, or for loop diagrams as
in the present case), the above sum leads to violation of gauge
invariance. The problem is alleviated either by choosing a
transverse polarization sum, or by introducing
ghost fields to get rid of the unphysical gluon polarizations while
keeping the simple polarization sum above.
The ghost contributions are shown in Fig.~\ref{Fig:tcggGhost}.

Fig.~{\ref{Fig:tcggGhost}} requires some explanation.
The ghosts couple only to a gluon, i.e. there is a $g$--ghost--ghost
coupling but there is no coupling to quarks. Therefore, there is a
ghost anti-ghost diagram for each diagram with a triple gluon coupling.
Furthermore, since the ghost is not its anti-particle there are six
diagrams with ghosts as in Fig.~{\ref{Fig:tcggGhost}} and there is no
statistical
factor after phase-space integration, unlike the case of two
final gluons where a statistical factor of $1/2$ is inserted following phase
space integration. Note also that there are no interference terms between
ghost and ghostless amplitudes.
\footnote{The ghost fields,
although they are bosons, obey
Fermi-Dirac statistics
and will have an overall -1 factor for each closed loop, similar to a fermion
loop.}

In our calculation we make the GIM mechanism manifest by dropping
   terms independent
     of internal quark mass.\footnote{This simplification applies to
the divergent parts
      as well since the SM without the unitarity of the CKM matrix is
non-renormalizable.}
      This will substantially simplify the calculation. The divergent
parts originate
    only  from the unphysical Higgs loops.
As the cancellation of ultraviolet divergences should happen  at the
amplitude squared
level for $t\to cgg$ in the 't Hooft - Feynman gauge, the analytical
output is so long that the
    ultraviolet finiteness has to be checked numerically.
We carried out the calculation in $D$-dimension
using dimensional
    regularization \cite{'tHooft:1973mm}
    and keep the decay width formulas as functions of
    $\epsilon = 4-D$. Then we check the stability of the decay width
   by varying
     $\epsilon$. Of course, this method is limited by the achievable
numerical precision
     of the program used. We have done our calculation with the
softwares FeynArts and FormCalc
     \cite{Hahn:2000jm, FF}
    and done some partial cross-checks with FeynCalc \cite{Mertig:1990wm}.
      With FormCalc the numerical check of ultraviolet finiteness is
controlled by
      a parameter and thus easier to check. We have tested and
confirmed the finiteness of our results for  $t\to cgg$
      (as well as for $t\to cq\bar{q},~q=u$ and $t\to cg$).

In addition to ultraviolet divergences, we must deal with infrared
and collinear divergences which exist in the calculation of $t \to cgg$. There
are three possible cases producing such singularities: a) the
configuration of having one of the gluons travel
parallel to the charm quark, b) the configuration
of having the gluons traveling parallel to each
other, or c) the fact that one of the gluons in
the final state could be soft. While cases a)
and b) are related collinear divergence,
case c) is related to the infrared singularity.
To cure case a) we take a non-zero mass
for the charm quark
in our numerical calculation even though we present
some of our analytical results in the $m_c=0$ limit, just for simplicity.

There are two ways to approach cases b) and c). One
can either do an extensive study by including QCD corrections
consistently of the same order in the perturbation theory, this case up
to the order of $\alpha_s^2$ requires including the
interference terms from two-loop $t\to cg$ (one-loop QCD corrections)
with one-loop $t \to cgg$; or one can exclude the part of the
phase space containing singular points by simply imposing cuts
on the kinematics of the process. The former method can be achieved
by  dimensionally regularizing the phase space integrals with
$D=4-\epsilon_{IR}$ and making the divergence manifest. Then the
cancellation of infrared singularities is guaranteed by the
Kinoshita-Lee-Nauenberg (KLN) theorem
\cite{Kinoshita:1962ur, Lee:1964is, Sterman:1976jh}.
Basically one has to carry out a
study for top quark decays similar to the
$b\to sg$  evaluation done by Greub and Liniger\cite{Greub:2000sy}.
We chose to proceed by cutting the ``dangerous'' integration limits
as discussed below. Our calculation is therefore more in the spirit of
\cite{Simma:1990nr}.

When we carry out integration over phase space by using
the momentum delta functions and azimuthal symmetry, we are left with
only two non-trivial integration over variables, say,
the energy of one of the gluons ($E_3\equiv k_3^0$) and the
energy of the $c$-quark ($E_c\equiv k_2^0$),
where the energies are in rest frame of the decaying top
quark. Then in ($E_c,E_3$)
space, the region of integration becomes a triangular shape.
The singularities discussed above lie at the
boundaries of the region, infrared singularities at the vertices
and collinear singularities along the boundary
lines. Thus we have to impose cuts on both $E_c$ and $E_3$. We will discuss
this point further later in the section. Note that since we
constrain our phase space, our result for ${\rm Br}(t\to cgg)$
should be seen as
approximate and more conservative.

Let us now present some analytical expressions.
   For the sake of simplicity of the presentation, the masses
$m_c, m_d$ and $m_s$ are assumed to be zero, though we
    included their contributions in our numerical study. From
Figs.~\ref{Fig:tcgg} and \ref{Fig:tcggGhost}, the
    amplitude for the decay can be written in a compact form as
\begin{eqnarray}
\displaystyle
A_{\rm self} &=&\frac{\alpha \alpha_s V_{tb} V_{cb}^*}{4 m_t^2
m_W^2 s_{23} \sin^2\!\theta_W t\, t_{12} R_1 R_4} \left[-F_2 m_t
t_{12} R_2 R_4-2 F_1 (SP_4-SP_5)
t R_6 R_4\right.\nn\\
&&\left.+s_{23} SP_1 R_1 (2 F_9 R_5-F_{12} m_t R_3 R_4)+F_3 t_{12}
R_7+2 F_4 R_8\right],\nn\\
A_{\rm vert} &=& \frac{\alpha  \alpha_s V_{tb} V_{cb}^*}{2 m_W^2
s_{23} \sin^2\!\theta_W t t_{12} R_1 R_4}
\left[2 R_1 (F_{13} m_t t\, t_{12} R(9)-F_9 s_{23} SP_1 R_{11})-2 F_4
R_{13}\right.\nn\\
&&\left.-F_3 t_{12} R_{14}+R_4
(m_t s_{23} t_{12} (2 F_{15} R_{10}+F_2 R_{12})+2 F_1 t
R_{15})+F_{12} m_t R_{16}\right],\nn\\
A_{\rm 1PI} &=& \frac{\alpha  \alpha_s V_{tb} V_{cb}^*}{2 m_W^2
\sin^2\!\theta_W}
\left[2 F_9 R_{17}+F_3 R_{19}+m_t \left(-2 F_{15} R_{18}+2 F_13
R_{20}+F_2 R_{21}\right.\right.\nn\\
&&\left.\left.-2 F_{12} R_{22}\right)+2 F_4 R_{23}-2 F_1 R_{24}\right],\nn\\
A_{\rm ghost} &=&\frac{\alpha  \alpha_s V_{tb} V_{cb}^*} {4 m_t^2
m_W^2 \sin^2\!\theta_W t_{12}} R_{25} \left[F_9 R_{27}-2 F_{12} m_t
R_{26}\right],\label{eq:Amplitude}
\end{eqnarray}
where
\begin{eqnarray}
\displaystyle
F_1 &=& \bar{u}(k_2,0) P_R \epsilon\!\!/^*(k_3)u(k_1,m_t),\;\;\;
F_2 = \bar{u}(k_2,0) P_R \epsilon\!\!/^*(k_3) \epsilon\!\!/^*(k_4)
u(k_1,m_t),\nn\\
F_3 &=& \bar{u}(k_2,0) P_R \epsilon\!\!/^*(k_3) \epsilon\!\!/^*(k_4)
k\!\!\!/_3 u(k_1,m_t),\;\;\;
F_4 = \bar{u}(k_2,0) P_R \epsilon\!\!/^*(k_4)u(k_1,m_t),\nn\\
F_5 &=&  \bar{u}(k_2,0) P_L \epsilon\!\!/^*(k_3) u(k_1,m_t),\;\;\;
F_6 = \bar{u}(k_2,0) P_L \epsilon\!\!/^*(k_3)
\epsilon\!\!/^*(k_4)u(k_1,m_t),\nn\\
F_7 &=& \bar{u}(k_2,0) P_L \epsilon\!\!/^*(k_3) \epsilon\!\!/^*(k_4)
k\!\!\!/_3 u(k_1,m_t),\;\;\;
F_8 =  \bar{u}(k_2,0) P_L \epsilon\!\!/^*(k_4) u(k_1,m_t),\nn\\
F_9 &=& \bar{u}(k_2,0) P_R k\!\!\!/_3 u(k_1,m_t),\;\;\;\;\;\;
F_{10} = \bar{u}(k_2,0) P_L u(k_1,m_t),\nn\\
F_{11} &=& \bar{u}(k_2,0) P_L k\!\!\!/_3 u(k_1,m_t),\;\;\;\;\;\;
F_{12} = \bar{u}(k_2,0) P_R u(k_1,m_t),\nn\\
F_{13} &=& \bar{u}(k_2,0) P_R \epsilon\!\!/^*(k_3) k\!\!\!/_3 u(k_1,m_t),\;\;\;
F_{14} = \bar{u}(k_2,0) P_L \epsilon\!\!/^*(k_3) k\!\!\!/_3 u(k_1,m_t),\nn\\
F_{15} &=& \bar{u}(k_2,0) P_R \epsilon\!\!/^*(k_4) k\!\!\!/_3 u(k_1,m_t),\;\;\;
F_{16} = \bar{u}(k_2,0) P_L \epsilon\!\!/^*(k_4) k\!\!\!/_3 u(k_1,m_t),
\end{eqnarray}
The Lorentz invariant
$t$ and $s$ are the usual Mandelstam
variables, while $t_{12}=(k_1-k_2)^2$ and $s_{23}=(k_2+k_3)^2$ are
the generalized
Mandelstam variables. The scalar products are defined as
\begin{eqnarray}
SP_1 &=& \epsilon^*(k_3)\cdot \epsilon^*(k_4),\;\;\;\;\;
SP_2 = \epsilon^*(k_3)\cdot k_1,\nn\\
SP_3 &=&  \epsilon^*(k_3)\cdot k_2,\;\;\;\;\;
SP_4 = \epsilon^*(k_4)\cdot k_1,\;\;\;
SP_5 =  \epsilon^*(k_4)\cdot k_2.
\end{eqnarray}
The functions $R_1,...,R_{27}$ are defined in terms of
Passarino-Veltman functions \cite{Passarino:1978jh}
and are given explicitly in the appendix.\footnote{The
result is expressed in terms of Passarino-Veltman functions
\cite{Passarino:1978jh} but the
reduction to the scalar function $A_0,B_0,C_0,$ and $D_0$ has not been
carried out.
This is indeed one advantage of using FormCalc which does not require
such reduction
to save substantial CPU time.}
The decay width
can be written as
\begin{eqnarray}
d\Gamma(t\to cgg) &=& \frac{1}{2m_t}\sum_{\rm{spins}}|{\cal M}|^2
d\Phi_3(k_1;k_2,k_3,
k_4)\nonumber\\
d\Phi_3(k_1;k_2,k_3,k_4) &=& \frac{d^3k_2}{(2\pi)^3 2k_2^0} \frac{d^3k_3}
{(2\pi)^3 2k_3^0} \frac{d^3k_4}{(2\pi)^3 2k_4^0} (2\pi)^4
\delta^{(4)}(k_1-k_2-k_3-k_4),
\end{eqnarray}
where $|{\cal M}|^2$ is straightforward to
calculate from the amplitude given in Eq.~(\ref{eq:Amplitude}). The
phase space $d\Phi_3$ can be expressed in terms of energies of the
third and fourth particles,
    chosen to be the gluon pair in the rest frame of top quark. In this
frame, one can take the production
    plane as the $x-z$ plane and choose the $c$ quark momentum along the
$z$-axis. After rearranging the
volume elements
     and
     carrying out angular and  momenta integrals, we implement the
phase space cuts discussed previously
\begin{eqnarray}
d\Phi_3(k_1;k_2,k_3,k_4) = \frac{1}{32\pi^3}\int_{(k_3^0)^{\rm min}}^{(k_3^0)
^{\rm max}}dk_3^0\int_{(k_2^0)^{\rm min}}^{(k_3^0)^{\rm max}}dk_2^0,
\end{eqnarray}
where the limits are
\begin{eqnarray}
(k_2^0)^{\rm min} &=& {\rm Max}\left[C m_t,\frac{\sigma-|{\bf
k}_3|}{2}\right],\nonumber\\
(k_2^0)^{\rm max} &=& \frac{\sigma+|{\bf
k}_3|}{2}(1-2C),\nonumber\\
(k_3^0)^{\rm min} &=& C m_t,\nonumber\\
(k_3^0)^{\rm max} &=& \frac{m_t}{2}(1-2C),
   \end{eqnarray}
    with $\sigma = m_t -k_3^0$ (recall that for simplicity
we have assumed $m_c = 0$). Here $C$ is our cutoff parameter, which we
initially
take as $C=0.001$ and then study the effect of increasing its value. To
calculate the
    branching ratio, we assume that $t\to bW$ is the dominant decay mode
of the top  quark
    and use $\Gamma(t\to bW) = 1.55$ GeV.
   For the numerical
analysis, we have used the parameters \cite{Fusaoka:1998vc}
given in Table~\ref{parameters}.
\begin{table}[h]
	\caption{The parameters used in the numerical
calculation.}\label{parameters}
\begin{center}
      \begin{tabular}{c c c c c c}
      \hline\hline
$\alpha_s(m_t)$ &\hspace*{1cm} $\alpha(m_t)$ &\hspace*{1cm}
$\sin\theta_W(m_t)$ &\hspace*{1cm} $m_c(m_t)$ &\hspace*{1cm}
$m_b(m_t)$ &\hspace*{1cm} $m_t(m_t)$\\
\hline
    0.106829  &\hspace*{1cm}   0.007544    &\hspace*{1cm}  0.22
&\hspace*{1cm}  0.63 GeV  &\hspace*{1cm}  2.85 GeV &\hspace*{1cm}
174.3 GeV  \\
\hline \hline
\end{tabular}
\end{center}
\end{table}

Our result is
\begin{eqnarray}
{\rm Br}(t\to cgg) &\equiv& \frac{\Gamma(t\to cgg)}{\Gamma(t\to bW)}\nn\\
&=& 1.02\times 10^{-9},
\end{eqnarray}
while  for the two-body decay $t\to cg$ we get
\begin{eqnarray}
{\rm Br}(t\to cg) = 5.73\times 10^{-12}.
\end{eqnarray}
for the same parameter set. At this point, a comment is needed.
For completeness, and to check our procedure,
we recalculated
the two-body decay $t\to cg$ and our numerical value is around one order of
magnitude smaller than that of the Ref. \cite{Eilam:1990zc}
(see Fig.~2 of \cite{Eilam:1990zc}).
The main source of such disprepancy lies in the value for bottom 
quark mass. The pole mass
$m_b=5$ GeV is used in Ref. \cite{Eilam:1990zc}, while we have used the running
bottom quark mass at $m_t$ scale, $m_b=2.85$ GeV. Since the branching ratio is
proportional to $m_b^4$ (due to GIM suppression), our results differ 
by one order of magnitude
($(2.85/5)^4\sim 0.1$). The original SM calculation of $t\to cg$ in Ref.
\cite{Eilam:1990zc} was later updated \cite{Aguilar-Saavedra:2002ns},
where the running bottom quark mass was
used and ${\rm Br}(t\to cg) = 4.6\times 10^{-12}$ was obtained.

There is however one cautionary remark about the
$t\to cgg$ decay. The branching ratios were calculated as function of the
cutoff (see below). One should calculate the rate for $t\to cgg$ by
doing a complete higher order calculation. QCD corrections should be at
most of the order of $10\%$, which is the order of magnitude of
QCD corrections in $t\to bW$.
One can view $C$ as a detector cutoff.
The cutoff dependence of the branching ratio of $t\to cgg$
is given in Table~\ref{Cvalues}. As seen from Table~\ref{Cvalues},
the branching ratio is sensitive, but not
significantly so, to the $C$ parameter.
\begin{table}[htb]
	\caption{The cutoff dependence of the branching ratio of $t\to cgg$ for
	various $C$ values.}
	\label{Cvalues}
\begin{center}
      \begin{tabular}{c c c c c}
      \hline\hline
$C$ (in $m_t$ units) &\hspace*{1cm} 0.001 &\hspace*{1cm} 0.003
&\hspace*{1cm} 0.005 &\hspace*{1cm} 0.01 \\
\hline
   ${\rm Br}(t\to cgg)$ &\hspace*{1cm}   $1.02\times10^{-9}$    &\hspace*{1cm}
   $9.04\times10^{-10}$  &\hspace*{1cm}  $8.76\times10^{-10}$  &\hspace*{1cm}
   $7.78\times10^{-10}$ \\
\hline \hline
\end{tabular}
\end{center}
\end{table}

In general, contributions from
ghost diagrams are quite suppressed with respect to the rest of the
diagrams and the 1PI diagrams slightly dominate the self energy and
vertex type diagrams depicted in Figs.~\ref{Fig:tcgg}.

Within the range $(0.001-0.01)$ for the cutoff $C$,
the rate for the three-body decay $t\to cgg$
is more than two orders of magnitude larger than the rate
for the two-body decay $t\to cg$. This is
higher order dominance par excellence.
However,
   the branching ratio for the $t\to cgg$ channel in the SM,
   is still too small to be
   observable in any conceivable experiment. Thus any experimental
    sighting of $t\to cgg$ will indicate the appearance of NP.
   The three-body decay $t\to cgg$ has another important
   difference with respect to $t \to cg$: A promising ratio for $t\to cgg$ might
   lead to a sizable cross section for single top production via $gg\to
   t\bar{c}$ at a future hadron collider, especially since most of
   the interesting events will be fed by partonic sub-processes
   originating from gluon-gluon collisions.
   We are currently investigating the decay $t\to cgg$ and
the effect of its crossed
   partonic sub-process $gg\to t{\bar c}$ on single top production
at the LHC, within the minimal supersymmetric SM \cite{Eilam:2006rb}.
\section{The decay $t \to c q{\bar q},~q=u$}

\begin{figure}[h]
\vspace*{-4.7in}
     \centerline{\epsfxsize 9in {\epsfbox{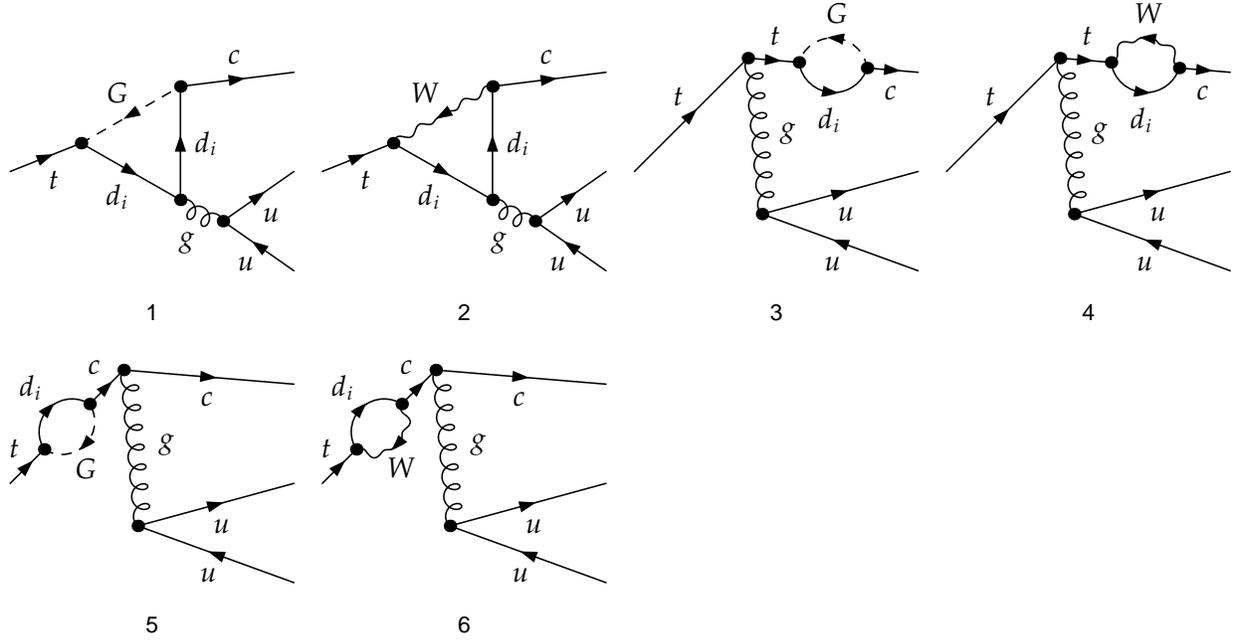}}}
\vspace*{-5.0in}
\caption{The dominant one-loop contributions to $t \to cq\bar{q}$ in the SM in
the 't Hooft -
Feynman gauge.} \label{fig:tcqqbarGstr}
\end{figure}

\begin{figure}[h]
\vspace*{-4.7in}
       \centerline{\epsfxsize 9in {\epsfbox{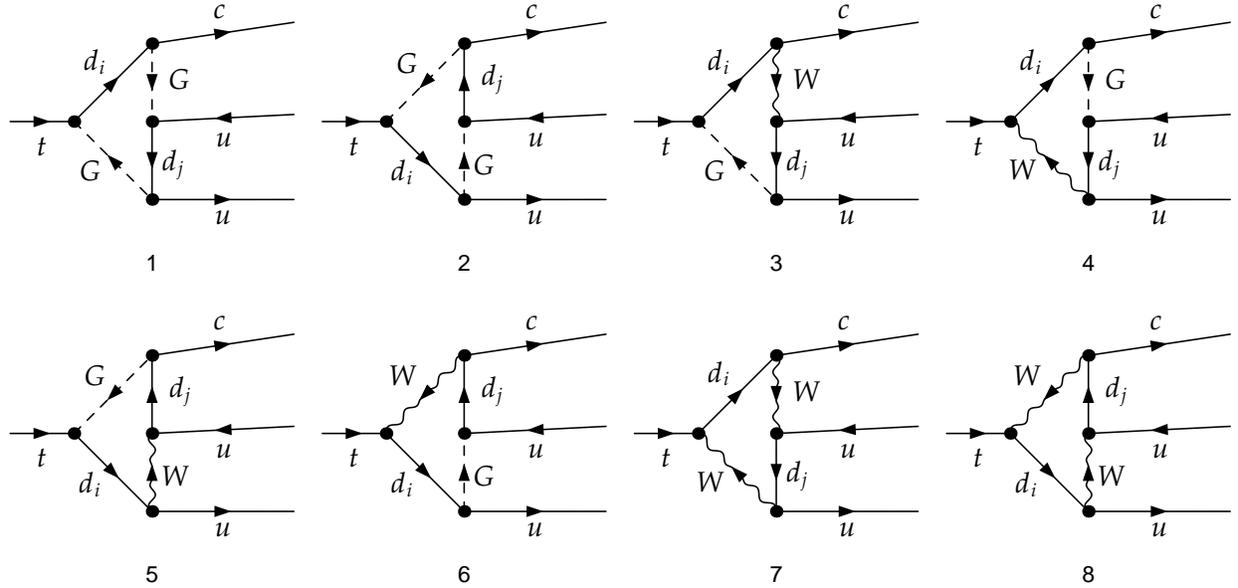}}}
\vspace*{-4.9in}
\caption{The sub-dominant one-loop contributions to $t \to cq\bar{q}$
in the SM in the 't Hooft -
Feynman gauge.}\label{fig:tcqqbar1PI}
\end{figure}

Using the same
procedure and the same parameters as for $t\to cgg$, we have also
calculated the branching ratio of $t \to c q{\bar
q},q=u$. Unlike $t\to cgg$ case, this decay arises dominantly
from the $tcg^{\star}$
vertex, where
$g^{\star}$ then decays to $\to q {\bar q}$ pair. There are of course diagrams
mediated
by electroweak  gauge bosons, $\gamma$ and $Z$  or the neutral Higgs boson,
but their contributions in the SM are
negligible. The dominant diagrams are given in
Fig.~\ref{fig:tcqqbarGstr}, while the
box diagrams displayed in Fig.~\ref{fig:tcqqbar1PI},
were found to be at least 3 orders of magnitude smaller than the diagrams
in Fig.~\ref{fig:tcqqbarGstr}.

Compared with $t\to cgg$, it is a much simpler calculation, mainly
since there are no external gluons and thus no ghosts.
The only non-trivial issue is to demonstrate the
ultraviolet finiteness of the decay. This can be checked either
analytically at the amplitude level, or numerically
for the branching ratio. We followed the second method.

We find that $t\to cq\bar{q},~q=u$ remains
slightly smaller than $t\to cg$. Thus
    we disagree with the recent study by Cordero-Cid et al.
\cite{Cordero-Cid:2004hk} where $t\to cq\bar{q},~q=u$ is found to be almost an
order of magnitude larger than $t\to cg$. The reason for the
discrepancy might be their use of different parameters, which
unfortunately are not spelled out in their paper.
Since we mainly concentrate on the $t\to cgg$ decay mode in this study, we
prefer to skip the details of this calculation and present only our result
here. For the parameter set chosen before, we found
\begin{eqnarray}
{\rm Br}(t\to cq\bar{q}, ~q=u)=3.96\times 10^{-12}.
\end{eqnarray}
As seen, unlike ${\rm Br}(t\to cgg)$, this ratio
stays slightly smaller but still comparable  to
${\rm Br}(t\to cg)$.
The decay $t\to c q\bar{q}$ has been analyzed
and compared with $t\to cg$ in both the SM and version II of the 
Two-Higgs-Doublet model
  by Deshpande, Margolis and Trottier \cite{Deshpande:1991pn}.\footnote{Some of this decay 
mode ($q=d,b$) has also been considered by Eilam et. al. in the SM \cite{Eilam:1991yv}.} They show that
for $m_t=175$ GeV, ${\rm BR}(t\to c q\bar{q})$ in the SM is slightly 
($\sim 1.2$ times) bigger than ${\rm Br}(t\to cg)$ when there is a
sum over $q=u,c,d,s,b$. Our results agree with theirs. We get $\sum 
\limits_{q=1}^{5}{\rm Br}(t\to cq\bar{q})$ as $1.56\times 10^{-11}$,
which becomes bigger than ${\rm Br}(t\to cg)$.

\section{Conclusions}\label{sec:conclusions}
In this study, we have discussed
higher order dominance in the rare top decays $t\to cgg$ within the
SM framework, a phenomenon known more than a decade ago in
$b$-physics.
For completeness, we also calculated $t\to cq\bar{q},~q=u$.
Using running quark masses, we have found
${\rm Br}(t\to cgg)= 1.02\times 10^{-9}$ with
a cutoff $C=0.001$ for a top mass
$m_t=174.3$ GeV. We considered the
sensitivity of the ratio with respect
to the cutoff parameter $C$ and found that even for
$C=0.01$ it is still more than
two orders of magnitude larger than the two-body decay ${\rm Br}(t\to cg)$
which we calculated as $5.73\times10^{-12}$ with the same set of
parameters. By comparison, we found
${\rm Br}(t\to c q {\bar
q},~q=u)$ to be smaller, $3.96 \times 10^{-12}$ and comparable to
${\rm Br}(t\to cg)$. However, when we sum over $q\bar{q}$ pairs
for all five quarks, it becomes slightly larger than ${\rm Br}(t\to cg)$.

If higher
order dominance is still valid for a viable NP model
in the sense that $t\to cg$  is much smaller than
$t\to cgg$ yet larger than its value in the SM,
we may have a glimpse of NP at work either in the
decay $t\to cgg$ or in production at the LHC through the
partonic sub-process $gg\to t{\bar c}$.

\begin{acknowledgments}
This work was funded by NSERC of Canada (SAP0105354).
The work of G.E. was supported in part by the Israel Science Foundation
and by the Fund for the Promotion of Research at the Technion. G.E.
also thanks N.G. Deshpande for helpful discussions.
\end{acknowledgments}

\appendix*
\section{The One-Loop R-functions}\label{sec:Appendix}
The explicit form of the functions defined in
Eq.~(\ref{eq:Amplitude}) are given below. As in the paper,
expressions are given for $m_c=0$, while we use a non-zero c-quark
mass in our numerical calculations.
We define
$T_d=(T_a T_b)_{ij}$, $T_e=(T_b T_a)_{ij}$ as products of color matrix
elements.  Here $a$ and $b$ are gluon indexes running from 1 to 8
with color indexes $i,j=1,2,3$
\begin{eqnarray}
R_1&=&m_t^2-s_{23},\nn\\
R_2&=&2 m_t^2 m_W^2 s_{23} (-t T_d+m_t^2 T_e-s_{23}
T_e)+((m_b^2-m_W^2) (m_b^2+2 m_W^2)
(m_t^2-s_{23}) s_{23} T_e\nn\\
&&-m_t^2 (m_b^4-2 m_W^4+m_b^2 (m_W^2-2 s_{23})) t T_d) B_0^{(1)}
-(m_b^4+2 (m_t^2-m_W^2) m_W^2\nn\\
&&+m_b^2 (m_W^2-m_t^2)) (m_t^2-s_{23}) s_{23} T_e B_0^{(2)}+2 m_b^2
m_t^2 s_{23} (s_{23}-m_t^2) T_e B_0^{(3)}\nn\\
&&+m_t^2 (m_b^4+m_b^2 (m_W^2-s_{23})+2 m_W^2(s_{23}-m_W^2)) t T_d
B_0^{(5)},\nn\\
R_3&=&(m_b^2-m_W^2) (m_b^2+2 m_W^2) (t T_d-(t+2 t_{12}) T_e)
(B_0^{(1)}-B_0^{(2)})
+m_t^2 (m_b^2 (t T_d (B_0^{(2)}\nn\\
&&-2 B_0^{(1)})+T_e (2 t B_0^{(1)}-t B_0^{(2)}-2 t_{12} B_0^{(2)}+4
t_{12} B_0^{(3)}))\nn\\
&&-2 m_W^2 (t T_d-(t+2 t_{12})T_e) (B_0^{(2)}-1)),\nn\\
R_4&=&m_t^2-t,\nn\\
R_5&=&2 m_t^2 m_W^2 ((m_t^2-t) t (T_d-T_e)-(m_t^2-2 t) t_{12} T_e)+(m_b^4
+m_b^2 (m_W^2-2 m_t^2)\nn\\
&&-2 m_W^4) t ((m_t^2-t) (T_d-T_e)+t_{12} T_e) B_0^{(1)}-(m_b^4+2 (m_t^2-m_W^2)
m_W^2\nn\\
&&+m_b^2 (m_W^2-m_t^2)) (m_t^2-t) (t T_d-(t+t_{12}) T_e) B_0^{(2)}+m_t^2
(-m_b^4-2 m_W^2 (t-m_W^2)\nn\\
&&-m_b^2 (-2 m_t^2+m_W^2+t)) t_{12} T_e B_0^{(3)},\nn\\
R_6&=&2 m_W^2 (m_t^2 (m_t^2-2 s_{23}) t_{12} T_d+m_t^2 (m_t^2-s_{23}) s_{23}
(T_d-T_e))+(m_b^4+m_b^2(m_W^2-2 m_t^2)\nn\\
&&-2 m_W^4) s_{23} ((m_t^2-s_{23}) (T_d-T_e)-t_{12}
T_d) B_0^{(1)}-(m_b^4+2 (m_t^2-m_W^2) m_W^2\nn\\
&&+m_b^2 (m_W^2-m_t^2)) (m_t^2-s_{23})
(t_{12}T_d+s_{23}(T_d-T_e))B_0^{(2)}+m_t^2
(m_b^4\nn\\
&&+2 m_W^2 (s_{23}-m_W^2)+m_b^2 (-2 m_t^2+m_W^2+s_{23})) t_{12} T_d
B_0^{(5)},\nn\\
R_7&=&2 m_t^2 m_W^2 (m_t^4 (t T_d+s_{23} T_e)+2 s_{23} t (t T_d+s_{23} T_e)
-m_t^2 (T_e s_{23}^2+2 t (T_d+T_e) s_{23}\nn\\
&&+t^2 T_d))-(m_b^4+m_b^2 (m_W^2-2 m_t^2)-2 m_W^4) s_{23} t (-t T_d-s_{23} T_e
+m_t^2 (T_d+T_e)) B_0^{(1)}\nn\\
&&-(m_b^4 +2(m_t^2-m_W^2) m_W^2 +m_b^2 (m_W^2-m_t^2)) (m_t^2-s_{23})
(m_t^2-t)(t T_d
+s_{23} T_e) B_0^{(2)}\nn\\
&&+m_t^2 (m_t^2 -s_{23}) s_{23} (m_b^4+2 m_W^2 (t-m_W^2)+m_b^2 (-2
m_t^2+m_W^2+t)) T_e B_0^{(3)}
+m_t^2(m_b^4\nn\\
&&+2 m_W^2 (s_{23}-m_W^2)+m_b^2 (-2 m_t^2+m_W^2+s_{23})) (m_t^2-t) t
T_d B_0^{(5)},\nn\\
R_8&=&2 m_t^2 m_W^2(s_{23} (s_{23}-m_t^2) (SP_2-SP_3) t (t-m_t^2)
(T_d-T_e)-t_{12}
(m_t^4 (SP_3 t T_d+s_{23}SP_2 T_e)\nn\\
&&+2 s_{23} t (SP_3 t T_d+s_{23} SP_2 T_e)
-m_t^2 (SP_3 t (2 s_{23}+t) T_d+s_{23} SP_2(s_{23}+2 t) T_e)))\nn\\
&&+(m_b^4+m_b^2 (m_W^2-2 m_t^2)-2 m_W^4) s_{23} t ((m_t^2-s_{23})
(SP_2-SP_3) (m_t^2-t)
(T_d-T_e)\nn\\
&&+t_{12} (m_t^2 SP_3 T_d-SP_3 t T_d+m_t^2 SP_2 T_e-s_{23} SP_2 T_e))
B_0^{(1)}-
(m_b^4+2 (m_t^2-m_W^2) m_W^2\nn\\
&&+m_b^2 (m_W^2-m_t^2)) (m_t^2-s_{23}) (m_t^2-t) (s_{23}
(SP_2-SP_3) t (T_d-T_e)-t_{12} (SP_3 t T_d\nn\\
&&+s_{23} SP_2 T_e)) B_0^{(2)}-m_t^2 (m_t^2-s_{23}) s_{23} SP_2 (m_b^4+2 m_W^2
(t-m_W^2)\nn\\
&&+m_b^2 (-2 m_t^2+m_W^2+t))t_{12} T_e B_0^{(3)}-m_t^2 (m_b^4+2 m_W^2
(s_{23}-m_W^2)\nn\\
&&+m_b^2 (-2 m_t^2+m_W^2+s_{23})) SP_3 (m_t^2-t)t t_{12} T_d B_0^{(5)},\nn\\
R_9&=&s_{23} SP_5 T_e (2 m_W^2 (C_1^{(1)}+C_{12}^{(1)})-m_b^2
(C_0^{(1)}+C_1^{(1)}-C_{12}^{(1)}))+SP_4 (m_t^2-t)\nn\\
&&\times T_d (2 m_W^2 (C_{12}^{(4)}+C_{22}^{(4)})+m_b^2 (C_0^{(4)}+C_{12}^{(4)}
+2 C_2^{(4)}+C_{22}^{(4)})),\nn\\
R_{10}&=&SP_3 t T_d (m_b^2 (C_0^{(3)}+C_1^{(3)}-C_{12}^{(3)})-2 m_W^2
(C_1^{(3)}+C_{12}^{(3)}))-(m_t^2-s_{23})\nn\\
&&\times SP_2 T_e (2 m_W^2 (C_{12}^{(2)}+C_{22}^{(2)})+m_b^2
(C_0^{(2)}+C_{12}^{(2)}
+2 C_2^{(2)}+C_{22}^{(2)})),\nn\\
R_{11}&=&m_W^2 (m_t^2-t) t (T_d-T_e)-(m_b^2+2 m_W^2) (m_t^2-t) t B_0^{(4)}
(T_d-T_e)\nn\\
&&+(m_t^2-t) t (((m_b^2-m_t^2-m_W^2) (m_b^2+2 m_W^2)+2 m_W^2 t_{12})
C_0^{(5)}+2 (m_b^2+2 m_W^2)
C_0^{(5)}\nn\\
&&+2 m_W^2 (t_{12}-m_t^2) C_1^{(5)}-(m_b^2 m_t^2+4 m_W^2 m_t^2-2 m_W^2
t_{12}) C_2^{(5)})
(T_d-T_e)\nn\\
&&-m_W^2 (m_t^2-2 t) t_{12} T_e+(m_b^2+2 m_W^2) (m_t^2-2 t) t_{12}
T_e B_0^{(3)}
+t t_{12} T_e (-m_b^2 m_t^2 C_0^{(1)}\nn\\
&&+2 (m_b^2+2 m_W^2) C_0^{(1)}+2 m_W^2 (m_t^2-s_{23}-t_{12}) C_1^{(1)})+
(m_t^2-t) t_{12} T_e (m_b^2 m_t^2 C_0^{(2)}\nn\\
&&-2 (m_b^2+2 m_W^2) C_0^{(2)}+(m_b^2 m_t^2-2 m_W^2 t) C_2^{(2)}),\nn\\
R_{12}&=&m_W^2 (-t T_d+m_t^2 T_e-s_{23} T_e)-(m_b^2+2 m_W^2) (m_t^2-s_{23}) T_e
B_0^{(3)}+(m_b^2+2 m_W^2)\nn\\
&&\times t T_d B_0^{(5)}+m_b^2 (m_t^2-s_{23}) t T_e C_0^{(1)}+t T_d
(m_b^2 s_{23} C_0^{(3)}-2 (m_b^2+2 m_W^2) C_0^{(3)}\nn\\
&&-2 m_W^2 s_{23} C_1^{(3)})-(m_t^2-s_{23}) T_e (m_b^2 ((m_t^2-t)
(C_0^{(2)}+C_2^{(2)})-2 C_0^{(2)})\nn\\
&&-4 m_W^2 C_0^{(2)})-(m_t^2-s_{23}) t
T_d (m_b^2(C_0^{(4)}+C_2^{(4)})-2 m_W^2 C_2^{(4)}),\nn\\
R_{13}&=&m_W^2 s_{23} (s_{23}-m_t^2) (SP_2-SP_3) t (t-m_t^2)
(T_d-T_e)-(m_b^2+2 m_W^2)
(m_t^2-s_{23})\nn\\
&&\times s_{23} (SP_2-SP_3) (m_t^2-t) t B_0^{(4)}
(T_d-T_e)+(m_t^2-s_{23}) s_{23} (SP_2-SP_3)
(m_t^2-t) t\nn\\
&&\times (((m_b^2-m_t^2-m_W^2) (m_b^2+2 m_W^2)+2 m_W^2 t_{12})
C_0^{(5)}+2 (m_b^2+2
m_W^2) C_0^{(5)}\nn\\
&&+2 m_W^2(t_{12}-m_t^2) C_1^{(5)}-(m_b^2 m_t^2+4 m_W^2 m_t^2-2 m_W^2
t_{12}) C_2^{(5)}) (T_d-T_e)
\nn\\
&&-m_W^2 t_{12} (m_t^4 (SP_3 t T_d+s_{23} SP_2 T_e)+2 s_{23} t (SP_3
t T_d+s_{23} SP_2 T_e)-m_t^2
(SP_3 t (2 s_{23}+t) T_d\nn\\
&&+s_{23} SP_2 (s_{23}+2 t) T_e))+(m_b^2+2 m_W^2) (m_t^2-s_{23})
s_{23} SP_2 (m_t^2-2 t) t_{12}
T_e B_0^{(3)}\nn\\
&&+(m_b^2+2 m_W^2) (m_t^2-2 s_{23}) SP_3 (m_t^2-t) t t_{12} T_d
B_0^{(5)}-(m_t^2
-s_{23}) s_{23} SP_2 t t_{12} T_e\nn\\
&&\times (m_b^2 m_t^2 C_0^{(1)}-2 (m_b^2+2 m_W^2) C_0^{(1)}+2
m_W^2(-m_t^2+s_{23}+t_{12}) C_1^{(1)})\nn\\
&&+s_{23} SP_3 (m_t^2-t) t t_{12} T_d (2 (m_b^2+2 m_W^2) C_0^{(3)}+m_b^2
m_t^2 C_1^{(3)}-(m_b^2+2 m_W^2)
s_{23} C_{12}^{(3)})\nn\\
&&+(m_t^2-s_{23}) SP_3 (m_t^2-t) t t_{12} T_d (m_b^2 m_t^2 C_0^{(4)}-2
(m_b^2+2 m_W^2) C_0^{(4)}
+(m_b^2 m_t^2\nn\\
&&+2 m_W^2 (-m_t^2+t+t_{12})) C_2^{(4)})-(m_t^2-s_{23}) s_{23} SP_2
(m_t^2-t) t_{12} T_e \nn\\
&&\times
(2 (m_b^2+2 m_W^2) C_0^{(2)}+(m_b^2+2 m_W^2)
(m_t^2-t) C_{12}^{(2)}+m_b^2 m_t^2 C_2^{(2)}\nn\\
&&+m_t^2 (m_b^2+2 m_W^2) C_{22}^{(2)}),\nn\\
R_{14}&=&m_W^2 (m_t^4 (t T_d+s_{23} T_e)+2 s_{23} t (t T_d+s_{23}
T_e)-m_t^2 (T_e
s_{23}^2\nn\\
&&+2 t(T_d+T_e) s_{23}+t^2 T_d))-(m_b^2+2 m_W^2) (m_t^2-s_{23})
s_{23} (m_t^2-2 t)
T_e B_0^{(3)}\nn\\
&&-(m_b^2+2 m_W^2) (m_t^2-2 s_{23}) (m_t^2-t) t T_d
B_0^{(5)}+(m_t^2-s_{23}) s_{23}
t T_e (m_b^2 m_t^2
C_0^{(1)}\nn\\
&&-2 (m_b^2+2 m_W^2) C_0^{(1)}+2 m_W^2 (-m_t^2+s_{23}+t_{12}) C_1^{(1)})+
s_{23} (m_t^2-t) t T_d (m_b^2 m_t^2 C_0^{(3)}\nn\\
&&-2 (m_b^2+2 m_W^2) C_0^{(3)}-2 m_W^2 s_{23} C_1^{(3)})-
(m_t^2-s_{23}) s_{23} (m_t^2-t) T_e (m_b^2 m_t^2 C_0^{(2)}\nn\\
&&-2 (m_b^2+2 m_W^2) C_0^{(2)}+(m_b^2 m_t^2-2
m_W^2 t) C_2^{(2)})-(m_t^2-s_{23}) (m_t^2-t) t T_d\nn\\
&&\times (m_b^2 m_t^2 C_0^{(4)}-2 (m_b^2+2 m_W^2) C_0^{(4)}+(m_b^2
m_t^2+2 m_W^2
(-m_t^2+t+t_{12})) C_2^{(4)}),\nn\\
R_{15}&=&m_W^2 (SP_4-SP_5) ((m_t^2-2 s_{23}) t_{12} T_d+(m_t^2-s_{23}) s_{23}
(T_d-T_e))-(m_b^2+2m_W^2)\nn\\
&&\times (m_t^2-s_{23}) s_{23} (SP_4-SP_5) (T_d-T_e) B_0^{(4)}-
(m_b^2+2 m_W^2) (m_t^2-2 s_{23}) (SP_4-SP_5) \nn\\
&&\times t_{12} T_d B_0^{(5)}+s_{23} (SP_4-SP_5) t_{12}
T_d (m_b^2 m_t^2 C_0^{(3)}-2 (m_b^2+2 m_W^2) C_0^{(3)}\nn\\
&&-2 m_W^2 s_{23} C_1^{(3)})-
(m_t^2-s_{23}) s_{23} SP_5 t_{12} T_e (m_b^2 C_{12}^{(1)}+2 m_W^2
(C_1^{(1)}+C_{12}^{(1)}))\nn\\
&&+(m_t^2-s_{23})
t_{12} T_d (-m_b^2 m_t^2 (SP_4-SP_5) C_0^{(4)}+(m_b^2+2 m_W^2) (2
(SP_4-SP_5) C_0^{(4)}\nn\\
&&-s_{23}
SP_4 C_{12}^{(4)})-(m_b^2 m_t^2 (SP_4-SP_5)+2 m_W^2 (m_t^2
(SP_5-SP_4)-SP_5 (t+t_{12})\nn\\
&&+
SP_4 (s_{23}+t+t_{12})))C_2^{(4)})+(m_t^2-s_{23}) s_{23} (SP_4-SP_5)
(T_d-T_e)\nn\\
&&\times (((m_b^2-m_t^2-m_W^2) (m_b^2+2 m_W^2)+2 m_W^2 t_{12})
C_0^{(5)}+2 (m_b^2
+2 m_W^2) C_0^{(5)}\nn\\
&&+2 m_W^2(t_{12}-m_t^2) C_1^{(5)}-(m_b^2 m_t^2+4 m_W^2 m_t^2-2 m_W^2 t_{12})
C_2^{(5)}),\nn\\
R_{16}&=&m_W^2 (m_t^2-s_{23}) s_{23} SP_1 (m_t^2-t) (t T_d-(t+2 t_{12}) T_e)+
2 (m_b^2+2 m_W^2)\nn\\
&&\times (m_t^2-s_{23}) s_{23} SP_1 (m_t^2-t) t_{12} T_e
B_0^{(3)}-(m_b^2+2 m_W^2)
    (m_t^2-s_{23})\nn\\
&&\times
s_{23} SP_1 (m_t^2-t) t (T_d-T_e) B_0^{(4)}-
2 (m_t^2-s_{23}) s_{23} t t_{12} T_e (m_b^2 (m_t^2 SP_1 C_0^{(1)}\nn\\
&&-SP_1 t C_0^{(1)}+2 SP_2 SP_5
(C_0^{(1)}+C_1^{(1)}-C_{12}^{(1)}))-4 m_W^2 SP_2 SP_5 (C_1^{(1)}\nn\\
&&+C_{12}^{(1)}))-
4 s_{23} SP_3 (SP_4-SP_5) (m_t^2-t) t t_{12} T_d (m_b^2
(C_0^{(3)}+C_1^{(3)}-C_{12}^{(3)})
\nn\\
&&-2 m_W^2 (C_1^{(3)}+C_{12}^{(3)}))+2 (m_t^2-s_{23}) s_{23} SP_1 (m_t^2-t)
t_{12} T_e (m_b^2 ((m_t^2-t) (C_0^{(2)}\nn\\
&&+C_2^{(2)})-2 C_0^{(2)})-4 m_W^2
C_0^{(2)})+4 (m_t^2-s_{23}) SP_3 SP_4 (m_t^2-t) t t_{12} T_d (2 m_W^2
(C_{12}^{(4)}\nn\\
&&+C_{22}^{(4)})
+m_b^2 (C_0^{(4)}+C_{12}^{(4)} +2C_2^{(4)}+C_{22}^{(4)}))+(m_t^2-s_{23}) s_{23}
(m_t^2-t) t (T_d-T_e)\nn\\
&&\times
(m_b^4 SP_1 C_0^{(5)}+2 m_W^2 (-m_W^2 SP_1 C_0^{(5)}-2s_{23} SP_1 C_0^{(5)}
-4 SP_2 SP_5 C_0^{(5)}+2 SP_1 C_0^{(5)}\nn\\
&&-2 s_{23} SP_1 C_1^{(5)}-4 SP_2 SP_5 C_1^{(5)}+
m_t^2 SP_1 C_{12}^{(5)}-2 s_{23} SP_1 C_{12}^{(5)}-4 SP_2 SP_5
C_{12}^{(5)}\nn\\
&&
-SP_1 t_{12} C_{12}^{(5)}-2 m_t^2
SP_1 C_2^{(5)}
-2 s_{23} SP_1 C_2^{(5)}-8 SP_2 SP_5 C_2^{(5)}
+2 SP_1 t C_2^{(5)}+SP_1 t_{12} C_2^{(5)}
\nn\\
&&-(4 SP_2 SP_5
+SP_1 (m_t^2-2 t-t_{12})) C_{22}^{(5)}+4 SP_3 SP_4 (C_0^{(5)}
+C_1^{(5)}+C_{12}^{(5)}+2 C_2^{(5)}\nn\\
&&+C_{22}^{(5)}))+m_b^2
(4 (SP_3 SP_4-SP_2 SP_5) (-C_1^{(5)}+C_{12}^{(5)}+C_{22}^{(5)})-
m_t^2 SP_1 (C_0^{(5)}+C_1^{(5)}\nn\\
&&-C_{12}^{(5)}+C_2^{(5)}+C_{22}^{(5)})
+SP_1 (m_W^2 C_0^{(5)}
+2 C_0^{(5)}+(2 s_{23}+t_{12}) (C_1^{(5)}-C_{12}^{(5)})\nn\\
&&+(2
t+t_{12}) C_{22}^{(5)}))),\nn\\
R_{17}&=&2 m_W^2 SP_1 T_d (C_0^{(6)}-C_1^{(6)})+m_b^2 SP_1 T_e
(C_2^{(7)}-C_0^{(7)})
+2 T_d (m_b^2 (SP_1 (D_0^{(1)}+D_2^{(1)}\nn\\
&&+m_W^2 (D_2^{(1)}-D_0^{(1)}))
+SP_3(SP_5 (D_{112}^{(1)}+D_{12}^{(1)}+D_{122}^{(1)})+SP_4
(D_{122}^{(1)}+D_{123}^{(1)}-D_{13}^{(1)}\nn\\
&&+D_{22}^{(1)}+D_{222}^{(1)}+D_{223}^{(1)}))+
SP_2 (SP_5 (D_{123}^{(1)}+D_{13}^{(1)})+SP_4 (D_{223}^{(1)}
+D_{23}^{(1)}+D_{233}^{(1)})))\nn\\
&&+m_W^2 (m_W^2 SP_1 (D_0^{(1)}-D_2^{(1)})+
2 (SP_3 (SP_5 (D_{112}^{(1)}
+D_{12}^{(1)}+D_{122}^{(1)})+SP_4 (2 D_{12}^{(1)}\nn\\
&&+D_{122}^{(1)}+D_{123}^{(1)}
+D_{13}^{(1)}+2 D_2^{(1)}
+3 D_{22}^{(1)}+D_{222}^{(1)}+D_{223}^{(1)}+2D_{23}^{(1)}))+SP_2
(SP_5 (D_{123}^{(1)}\nn\\
&&-D_{13}^{(1)})+SP_4
(D_{223}^{(1)}+D_{23}^{(1)}+D_{233}^{(1)})))+SP_1 (-4 D_0^{(1)}
+2 D_2^{(1)}-s_{23} (D_{12}^{(1)}+2 D_{22}^{(1)})\nn\\
&&+(-2m_t^2+t+t_{12}) D_{23}^{(1)}
+m_t^2 D_3^{(1)})))+
T_e (m_b^4 SP_1 (D_0^{(2)}-D_3^{(2)})
-m_b^2 (m_t^2 SP_1\nn\\
&&\times (D_0^{(2)}-D_{13}^{(2)}+D_2^{(2)}
-D_{23}^{(2)})+SP_1 (-4 D_0^{(2)}+2 D_3^{(2)}+(s_{23}+t_{12})
D_{13}^{(2)}+m_W^2 (D_0^{(2)}\nn\\
&&-D_3^{(2)})-t (D_{23}^{(2)}+2 D_{33}^{(2)}))+
2 (SP_3 (SP_4 (D_{12}^{(2)}+D_{123}^{(2)})
+SP_5 (D_{113}^{(2)}+D_{13}^{(2)}\nn\\
&&+D_{133}^{(2)}))+
SP_2 (SP_4 (D_{223}^{(2)}+D_{23}^{(2)}+D_{233}^{(2)})+SP_5
(-D_{12}^{(2)}+D_{123}^{(2)}
+D_{133}^{(2)}+D_{233}^{(2)}+D_{33}^{(2)}\nn\\
&&+D_{333}^{(2)}))))-
2 m_W^2 (m_t^2 SP_1 (D_0^{(2)}+D_1^{(2)}+D_2^{(2)}+D_3^{(2)})-SP_1
(t_{12} (D_0^{(2)}+D_1^{(2)}
\nn\\
&&+D_2^{(2)}+D_3^{(2)})-2 (D_0^{(2)}+D_3^{(2)}))+2 (SP_3 (SP_4
(D_{123}^{(2)}-D_{12}^{(2)})
+SP_5 (D_{113}^{(2)}
+D_{13}^{(2)}\nn\\
&&+D_{133}^{(2)}))+
SP_2 (SP_4 (D_{223}^{(2)}+D_{23}^{(2)}+D_{233}^{(2)})
+SP_5 (D_{12}^{(2)}+D_{123}^{(2)}+2 D_{13}^{(2)}\nn\\
&&+D_{133}^{(2)}+2 D_{23}^{(2)}
+D_{233}^{(2)}+2D_3^{(2)}+3 D_{33}^{(2)}+D_{333}^{(2)}))))),\nn\\
R_{18}&=&m_b^2 (SP_3 (T_e (D_{12}^{(2)}-D_1^{(2)})+T_d
(-D_1^{(1)}+D_{12}^{(1)}+D_{13}^{(1)}
+D_{22}^{(1)}+D_{23}^{(1)}))\nn\\
&&+SP_2 (T_e(D_{22}^{(2)}+D_{23}^{(2)}-D_3^{(2)})+T_d
(D_{23}^{(1)}+D_{33}^{(1)})))
+2 m_W^2 (SP_3 (T_d
(D_0^{(1)}+D_1^{(1)}\nn\\
&&-D_{12}^{(1)}-D_{13}^{(1)}-D_{22}^{(1)}-D_{23}^{(1)}+D_3^{(1)})
+T_e (D_0^{(2)}+D_1^{(2)}-D_{12}^{(2)}+D_2^{(2)}+D_3^{(2)}))\nn\\
&&-
SP_2 (T_e (D_2^{(2)}+D_{22}^{(2)}
+D_{23}^{(2)})+T_d (D_{23}^{(1)}+D_3^{(1)}+D_{33}^{(1)}))),\nn\\
R_{19}&=&m_b^4 (T_d (D_2^{(1)}-D_0^{(1)})+T_e (D_3^{(2)}-D_0^{(2)}))+
m_b^2 (T_d (C_0^{(6)}-C_1^{(6)}-6 D_0^{(1)}
+3 m_W^2 (D_0^{(1)}\nn\\
&&-D_2^{(1)})-s_{23} (D_{12}^{(1)}+2 D_{22}^{(1)})
+(t+t_{12}) D_{23}^{(1)}+m_t^2 (D_0^{(1)}
-2D_{23}^{(1)}
+D_3^{(1)}))\nn\\
&&+T_e (C_0^{(7)}-C_2^{(7)}
-6 D_0^{(2)}+(s_{23}+t_{12}) D_{13}^{(2)}+m_t^2 (D_0^{(2)}
-D_{13}^{(2)}\nn\\
&&+D_2^{(2)}-D_{23}^{(2)})+3 m_W^2 (D_0^{(2)}-D_3^{(2)})-t (D_{23}^{(2)}
+2D_{33}^{(2)})))
+
2 m_W^2 (T_d \nn\\
&&\times(-C_0^{(6)}+C_1^{(6)}+6 D_0^{(1)}+m_W^2 (D_2^{(1)}-D_0^{(1)})
+s_{23} (D_{12}^{(1)}
+2 D_{22}^{(1)})\nn\\
&&-t D_{23}^{(1)}+m_t^2 (D_0^{(1)}+D_1^{(1)}+D_2^{(1)}+2D_{23}^{(1)})-
t_{12} (D_0^{(1)}
+D_1^{(1)}
+D_2^{(1)}+D_{23}^{(1)}\nn\\
&&+D_3^{(1)}))
+T_e (-C_0^{(7)}+C_2^{(7)}+6 D_0^{(2)}
-s_{23} D_{13}^{(2)}
+m_W^2 (D_3^{(2)}
-D_0^{(2)})-
t_{12} (D_0^{(2)}+D_1^{(2)}\nn\\
&&+D_{13}^{(2)}+D_2^{(2)}+D_3^{(2)})
+m_t^2 (D_0^{(2)}
+D_1^{(2)}+D_{13}^{(2)}+D_{23}^{(2)}+D_3^{(2)})+t (D_{23}^{(2)}+2
D_{33}^{(2)}))),\nn\\
R_{20}&=&2 m_W^2 (SP_5 (T_d
(D_0^{(1)}+D_1^{(1)}-D_{13}^{(1)}+D_2^{(1)}+D_3^{(1)})
+T_e (D_0^{(2)}
+D_1^{(2)}-D_{12}^{(2)}-D_{13}^{(2)}\nn\\
&&+D_2^{(2)}-D_{23}^{(2)}-D_{33}^{(2)}))-
SP_4 (T_e (D_2^{(2)}\
+D_{22}^{(2)}+D_{23}^{(2)})+T_d (D_{23}^{(1)}
+D_3^{(1)}+D_{33}^{(1)})))\nn\\
&&+
m_b^2 (SP_4 (T_e (D_{22}^{(2)}
+D_{23}^{(2)})+T_d (-D_2^{(1)}+D_{23}^{(1)}
+D_{33}^{(1)}))
+SP_5 (T_d (D_{13}^{(1)}-D_1^{(1)})\nn\\
&&
+T_e (-D_1^{(2)}
+D_{12}^{(2)}+D_{13}^{(2)}+D_{23}^{(2)}+D_{33}^{(2)}))),\nn\\
R_{21}&=&T_d (2 m_W^2 (C_0^{(6)}+C_2^{(6)})-m_b^2 C_2^{(6)})
+T_e ((m_b^2-2 m_W^2) (C_1^{(7)}
+C_2^{(7)})-m_b^2 C_0^{(7)})\nn\\
&&+
T_d (m_b^4 D_3^{(1)}+2 m_W^2 (-2 D_0^{(1)}-s_{23} (D_0^{(1)}
+D_1^{(1)}-D_{13}^{(1)}-2 D_{23}^{(1)})
+m_t^2 D_3^{(1)}\nn\\
&&+m_W^2 (D_0^{(1)}+D_3^{(1)})
-(-2m_t^2+t+t_{12}) D_{33}^{(1)})+
m_b^2 (2 D_0^{(1)}+s_{23} (D_1^{(1)}-D_{13}^{(1)}+D_2^{(1)}\nn\\
&&-2 D_{23}^{(1)})-m_W^2 (2 D_0^{(1)}+3 D_3^{(1)})+(-2
m_t^2+t+t_{12}) D_{33}^{(1)}))
+
T_e (m_b^4 (D_0^{(2)}-D_2^{(2)}\nn\\
&&-D_3^{(2)})+m_b^2 (4 D_0^{(2)}+(s_{23}
+t_{12}) (D_1^{(2)}
-D_{12}^{(2)}-D_{13}^{(2)})+m_t^2 (-D_0^{(2)}-D_1^{(2)}+D_{12}^{(2)}
\nn\\
&&+D_{13}^{(2)}-D_2^{(2)}
+D_{22}^{(2)}+D_{23}^{(2)})+
m_W^2 (3 (D_2^{(2)}+D_3^{(2)})-D_0^{(2)})+t (D_{22}^{(2)}
+3 D_{23}^{(2)}-D_3^{(2)}\nn\\
&&+2 D_{33}^{(2)}))
-2 m_W^2 (4 D_0^{(2)}-t_{12} (D_{12}^{(2)}
+D_{13}^{(2)})
+m_t^2(D_{12}^{(2)}+D_{13}^{(2)}+D_{22}^{(2)}+D_{23}^{(2)})
\nn\\
&&+m_W^2 (D_2^{(2)}+D_3^{(2)})
+s_{23} (D_0^{(2)}
+D_1^{(2)}-D_{12}^{(2)}-D_{13}^{(2)}+D_2^{(2)}+D_3^{(2)})
+t (D_2^{(2)}+D_{22}^{(2)}\nn\\
&&
+3 D_{23}^{(2)}+D_3^{(2)}+2 D_{33}^{(2)}))),\nn\\
R_{22}&=&2 m_W^4 SP_1 T_d (D_0^{(1)}+D_3^{(1)})+
m_b^2 (-SP_1 (2 T_d (D_3^{(1)}-D_0^{(1)})
+T_e (C_0^{(7)}-C_1^{(7)}-C_2^{(7)}\nn\\
&&+2 (-2 D_0^{(2)}+D_2^{(2)}+D_3^{(2)})-(s_{23}+t_{12})
(D_1^{(2)}
-D_{12}^{(2)}-D_{13}^{(2)})+m_t^2(D_0^{(2)}+
D_1^{(2)}\nn\\
&&-D_{12}^{(2)}-D_{13}^{(2)}+D_2^{(2)}-D_{22}^{(2)}
-D_{23}^{(2)})+m_b^2 (-D_0^{(2)}+D_2^{(2)}+D_3^{(2)})-t (D_{22}^{(2)}\nn\\
&&+3 D_{23}^{(2)}
-D_3^{(2)}
+2 D_{33}^{(2)})))-
2 (SP_3 (SP_5 (T_e (-D_{11}^{(2)}+D_{112}^{(2)}+D_{113}^{(2)}+D_{123}^{(2)}
+D_{133}^{(2)})\nn\\
&&-T_d (D_{11}^{(1)}-D_{113}^{(1)}+2
D_{12}^{(1)}-D_{123}^{(1)}+D_{22}^{(1)}+D_{23}^{(1)}))+
SP_4 (T_e (D_{122}^{(2)}
+D_{123}^{(2)})\nn\\
&&+T_d (-D_1^{(1)}
+D_{123}^{(1)}+D_{13}^{(1)}+D_{133}^{(1)}-D_2^{(1)}
+D_{223}^{(1)}
+2 D_{23}^{(1)}+D_{233}^{(1)}+D_{33}^{(1)})))\nn\\
&&+
SP_2 (SP_4 (T_e (D_{22}^{(2)}+D_{222}^{(2)}
+2 D_{223}^{(2)}+D_{23}^{(2)}+D_{233}^{(2)})+T_d
(D_{233}^{(1)}+D_{33}^{(1)}+D_{333}^{(1)}))\nn\\
&&+
SP_5 (T_e (-D_1^{(2)}+D_{122}^{(2)}+2
D_{123}^{(2)}+D_{133}^{(2)}+D_{223}^{(2)}+D_{23}^{(2)}
+2 D_{233}^{(2)}+D_{33}^{(2)}+D_{333}^{(2)})\nn\\
&&-T_d
(D_{13}^{(1)}-D_{133}^{(1)}+D_{23}^{(1)}+D_{33}^{(1)})))))
+
m_W^2 (SP_1 (2 T_d (C_0^{(6)}+C_2^{(6)}-2 D_0^{(1)}-2 D_3^{(1)}\nn\\
&&+s_{23} (D_{13}^{(1)}
+D_2^{(1)}
+2 D_{23}^{(1)})-m_b^2 (D_0^{(1)}+D_3^{(1)})+(2 m_t^2-t-t_{12})
(D_3^{(1)}+D_{33}^{(1)}))\nn\\
&&-
T_e (4 (D_0^{(2)}+D_2^{(2)}+D_3^{(2)})+m_b^2
(D_0^{(2)}-D_2^{(2)}-D_3^{(2)})+2 s_{23}
(D_0^{(2)}
+D_1^{(2)}+D_2^{(2)}\nn\\
&&+D_3^{(2)})))-
4 (SP_3 (SP_5 (T_e (D_1^{(2)}+D_{11}^{(2)}+D_{112}^{(2)}
+D_{113}^{(2)}+2 D_{12}^{(2)}+D_{123}^{(2)}
+2 D_{13}^{(2)}\nn\\
&&+D_{133}^{(2)})
+
T_d (D_1^{(1)}
+D_{11}^{(1)}+D_{113}^{(1)}+2 D_{12}^{(1)}+D_{123}^{(1)}+2
D_{13}^{(1)}+D_2^{(1)}+D_{22}^{(1)}
+D_{23}^{(1)}))\nn\\
&&+
SP_4 (T_e (D_{122}^{(2)}+D_{123}^{(2)})+T_d (D_0^{(1)}+D_1^{(1)}+D_{123}^{(1)}
+D_{13}^{(1)}+D_{133}^{(1)}+D_2^{(1)}+D_{223}^{(1)}\nn\\
&&+2 D_{23}^{(1)}+D_{233}^{(1)}
+2 D_3^{(1)}+D_{33}^{(1)})))
+
SP_2 (SP_4 (T_e (D_{22}^{(2)}+D_{222}^{(2)}+2
D_{223}^{(2)}+D_{23}^{(2)}+D_{233}^{(2)})\nn\\
&&+T_d (D_{233}^{(1)}+D_{33}^{(1)}+D_{333}^{(1)}))+
SP_5 (T_d (D_{13}^{(1)}+D_{133}^{(1)}+D_{23}^{(1)}
+D_3^{(1)}+D_{33}^{(1)})+T_e (D_0^{(2)}\nn\\
&&+D_1^{(2)}+2 D_{12}^{(2)}+D_{122}^{(2)}+2
D_{123}^{(2)}\
+
2 D_{13}^{(2)}+D_{133}^{(2)}+3 D_2^{(2)}
+2 D_{22}^{(2)}+D_{223}^{(2)}\nn\\
&&+5 D_{23}^{(2)}+2
D_{233}^{(2)}
+3 (D_3^{(2)}+D_{33}^{(2)})+D_{333}^{(2)}))))),\nn\\
R_{23}&=&2 m_W^4 (SP_2 (T_d D_0^{(1)}-T_e (D_0^{(2)}+D_2^{(2)}+D_3^{(2)}))
-SP_3 (T_d (D_0^{(1)}
+D_1^{(1)}+D_2^{(1)})\nn\\
&&-T_e D_0^{(2)}))
+
m_b^2 (-SP_2 T_e C_0^{(7)}-SP_3 (T_d (C_0^{(6)}
-m_b^2 D_0^{(1)}-4 D_0^{(1)}-2 D_1^{(1)}-2 D_2^{(1)}\nn\\
&&-s_{23} (D_{12}^{(1)}
+D_{22}^{(1)})
+(t+t_{12})
(D_{13}^{(1)}+D_{23}^{(1)})+
m_t^2 (D_0^{(1)}+D_1^{(1)}-D_{13}^{(1)}+D_2^{(1)}-D_{23}^{(1)}
\nn\\
&&+D_3^{(1)}))
-T_e (C_0^{(7)}+C_1^{(7)}+C_2^{(7)}+2 D_1^{(2)}+(m_b^2-m_t^2) D_1^{(2)}-t
(D_{12}^{(2)}
+D_{13}^{(2)})))\nn\\
&&-
SP_2 (-T_e (m_b^2 D_0^{(2)}
+2 (2 D_0^{(2)}+D_2^{(2)}
+D_3^{(2)})-(s_{23}+t_{12}) D_{12}^{(2)}+m_t^2 (-D_0^{(2)}\nn\\
&&+D_{12}^{(2)}+D_{13}^{(2)}))
+T_d
(C_2^{(6)}-2 D_3^{(1)}+s_{23} (D_{13}^{(1)}+D_{23}^{(1)})-
m_b^2 D_3^{(1)}+m_t^2 (D_3^{(1)}
+D_{33}^{(1)}))\nn\\
&&+T_e ((s_{23}+t_{12}) D_{13}^{(2)}+m_t^2
(D_2^{(2)}-D_{22}^{(2)}-D_{23}^{(2)}
+D_3^{(2)})-t (D_{23}^{(2)}+D_{33}^{(2)}))))\nn\\
&&+
m_W^2 (SP_3 (T_d (2 C_2^{(6)}+m_b^2 (D_0^{(1)}
+2 (D_1^{(1)}+D_2^{(1)}))+2 (2 (D_1^{(1)}+D_2^{(1)})-s_{23} (D_{12}^{(1)}\nn\\
&&+D_2^{(1)}
+2 D_{22}^{(1)})+t (D_{12}^{(1)}+D_{13}^{(1)}+D_2^{(1)}+D_{23}^{(1)})+
t_{12} (D_0^{(1)}
+2 D_1^{(1)}+D_{11}^{(1)}+2 D_{12}^{(1)}\nn\\
&&+D_{13}^{(1)}+2 D_2^{(1)}
+D_{23}^{(1)}+D_3^{(1)})
-
m_t^2 (D_0^{(1)}+2 D_1^{(1)}+D_{11}^{(1)}+3 D_{12}^{(1)}+3 D_{13}^{(1)}+3
D_2^{(1)}\nn\\
&&+D_{22}^{(1)}
+3 D_{23}^{(1)}+2 D_3^{(1)})))+T_e (2 C_0^{(7)}-m_b^2 (2 D_0^{(2)}+D_1^{(2)})
+
2 (-2 D_0^{(2)}+2 D_1^{(2)}\nn\\
&&+s_{23} D_{13}^{(2)}
+(t+t_{12}) (D_{12}^{(2)}+D_{13}^{(2)})
+t D_3^{(2)}+m_t^2 (D_0^{(2)}+D_1^{(2)}-D_{12}^{(2)}-D_{13}^{(2)}+2
D_2^{(2)}\nn\\
&&+D_3^{(2)}))))
+
SP_2 (-2 T_e (C_0^{(7)}+C_1^{(7)}+C_2^{(7)})+T_d (2 C_0^{(6)}-m_b^2
(2 D_0^{(1)}+D_3^{(1)})\nn\\
&&+2 (-2 D_0^{(1)}+2 D_3^{(1)}+s_{23} D_{13}^{(1)}+t_{12} D_{13}^{(1)}+
s_{23} D_2^{(1)}
+(s_{23}+t+t_{12}) D_{23}^{(1)}-m_t^2 (D_{13}^{(1)}\nn\\
&&+2 D_{23}^{(1)}
+D_{33}^{(1)})))
+T_e (m_b^2 (D_0^{(2)}+2 (D_2^{(2)}+D_3^{(2)}))+2 (2
(D_2^{(2)}+D_3^{(2)})+s_{23}
(D_{12}^{(2)}\nn\\
&&+D_{13}^{(2)}+D_{23}^{(2)}+D_3^{(2)})-m_t^2
(D_0^{(2)}+D_1^{(2)}+D_{12}^{(2)}+D_{13}^{(2)}+3
D_2^{(2)}
+
2 D_{22}^{(2)}+3 D_{23}^{(2)}\nn\\
&&+2 D_3^{(2)})+t_{12}
(D_0^{(2)}+D_1^{(2)}+D_{12}^{(2)}+D_{13}^{(2)}
+2 D_2^{(2)}+D_{22}^{(2)}
+2 (D_{23}^{(2)}+D_3^{(2)}))\nn\\
&&-t (D_{23}^{(2)}+D_3^{(2)}+2
D_{33}^{(2)}))))),\nn\\
R_{24}&=&-m_b^4 (SP_4 (T_d D_0^{(1)}+T_e D_2^{(2)})+SP_5 (T_d
(-D_0^{(1)}+D_1^{(1)}+D_2^{(1)})
+T_e D_3^{(2)}))\nn\\
&&+
2 m_W^2 (SP_5 (T_e (C_0^{(7)}-C_1^{(7)}-C_2^{(7)}-2 (3 D_0^{(2)}+D_1^{(2)}
+D_3^{(2)})+m_W^2 (2 D_0^{(2)}+D_1^{(2)})\nn\\
&&-t_{12} (D_1^{(2)}+D_{11}^{(2)})+s_{23}
D_{13}^{(2)}
+
m_t^2 (D_0^{(2)}+2 D_1^{(2)}+D_{11}^{(2)}+D_{12}^{(2)}+3
D_2^{(2)}+D_3^{(2)})\nn\\
&&+t (D_{12}^{(2)}
+2 (D_{13}^{(2)}+D_3^{(2)})))-T_d (C_1^{(6)}+4 D_0^{(1)}+2 D_1^{(1)}+m_W^2
D_2^{(1)}
+s_{23}
(D_{12}^{(1)}+D_2^{(1)}\nn\\
&&+2 D_{22}^{(1)})-t (D_{13}^{(1)}+D_{23}^{(1)})-t_{12}
(D_0^{(1)}+D_1^{(1)}
+D_2^{(1)}+D_{23}^{(1)}+D_3^{(1)})+m_t^2 (D_0^{(1)}\nn\\
&&+D_1^{(1)}+D_{13}^{(1)}+D_2^{(1)}+2
D_{23}^{(1)}
+D_3^{(1)})))+
SP_4 (T_d (C_0^{(6)}+C_1^{(6)}+C_2^{(6)}-2 D_2^{(1)}\nn\\
&&-2 D_3^{(1)}+s_{23} (D_{12}^{(1)}
+D_{13}^{(1)}+2 (D_2^{(1)}+D_{22}^{(1)}+D_{23}^{(1)}))+m_W^2
(D_0^{(1)}+D_2^{(1)}+D_3^{(1)})\nn\\
&&+
m_t^2 (D_0^{(1)}+D_1^{(1)}+D_2^{(1)}+D_{23}^{(1)}+2
D_3^{(1)}+D_{33}^{(1)})-t_{12} (D_0^{(1)}
+D_1^{(1)}+D_2^{(1)}\nn\\
&&+D_{23}^{(1)}+2 D_3^{(1)}+D_{33}^{(1)}))-
T_e (C_0^{(7)}+m_W^2 D_0^{(2)}
-2 D_0^{(2)}+2 D_2^{(2)}-s_{23} D_{12}^{(2)}+m_t^2 D_2^{(2)}\nn\\
&&+t D_3^{(2)})))
+
m_b^2 (SP_4 (T_e (C_1^{(7)}-2 D_2^{(2)}-(s_{23}+t_{12}) D_{12}^{(2)}+m_W^2 (2
D_0^{(2)}\nn\\
&&+D_2^{(2)})+t D_{23}^{(2)}+m_t^2 (D_{12}^{(2)}+D_2^{(2)}+2
D_{22}^{(2)}+D_{23}^{(2)}))
+
T_d (C_0^{(6)}-4 D_0^{(1)}-2 D_2^{(1)}\nn\\
&&-2 D_3^{(1)}-s_{23}
(D_{12}^{(1)}+D_{13}^{(1)}+D_{22}^{(1)}
+D_{23}^{(1)})+m_t^2 (D_0^{(1)}+D_3^{(1)})-m_W^2 (D_0^{(1)}+2 (D_2^{(1)}\nn\\
&&+D_3^{(1)}))))
+
SP_5 (T_e (C_2^{(7)}+2 D_0^{(2)}-2 D_1^{(2)}-2 D_3^{(2)}+m_W^2 (-4 D_0^{(2)}-2
D_1^{(2)}+D_3^{(2)})\nn\\
&&+t (D_{33}^{(2)}-D_{12}^{(2)})+m_t^2 (D_{12}^{(2)}+D_{13}^{(2)}+2
D_{23}^{(2)}
+D_3^{(2)}+D_{33}^{(2)}))-T_d (2 C_0^{(6)}+
C_2^{(6)}-6 D_0^{(1)}\nn\\
&&+2 D_1^{(1)}+m_W^2 (D_0^{(1)}
-D_1^{(1)}-3 D_2^{(1)})-2 s_{23}
(D_{12}^{(1)}+D_{22}^{(1)})+(t+t_{12}) (D_{13}^{(1)}\nn\\
&&+D_{23}^{(1)})+m_t^2(D_0^{(1)}-2
(D_{13}^{(1)}+D_{23}^{(1)})+D_3^{(1)})))),\nn\\
R_{25}&=&T_d-T_e,\nn\\
R_{26}&=&m_b^2 (s_{23} (C_{12}^{(5)}-C_1^{(5)})+(m_t^2-t) C_{22}^{(5)})
+2 m_W^2 (s_{23} (C_0^{(5)}+C_1^{(5)}+C_{12}^{(5)}
+C_2^{(5)})\nn\\
&&+(m_t^2-t)
(C_2^{(5)}+C_{22}^{(5)})),\nn\\
R_{27}&=&(m_b^4+m_b^2 (m_W^2-2 m_t^2)-2 m_W^4) B_0^{(1)}
-(m_b^4+2 (m_t^2-m_W^2) m_W^2
+m_b^2
(m_W^2-m_t^2)) B_0^{(2)}\nn\\
&&+2 m_t^2 (m_b^2+2 m_W^2) B_0^{(4)}+
2 m_t^2 (-((m_b^2-m_t^2-m_W^2)(m_b^2+2 m_W^2)+2 m_W^2 t_{12}) C_0^{(5)}\nn\\
&&-2 (m_b^2+2 m_W^2)
C_0^{(5)}+2 m_W^2 (m_t^2-t_{12}) C_1^{(5)}
+(m_b^2 m_t^2+4 m_W^2 m_t^2-2 m_W^2 t_{12}) C_2^{(5)})\nn.
\end{eqnarray}
where
\begin{eqnarray}
B_0^{(1)} &=& B_0(m_b^2,m_W^2),\;\;\;
B_0^{(2)} = B_0(m_t^2,m_b^2,m_W^2),\;\;\;
B_0^{(3)} = B_0(t,m_b^2,m_W^2),\nn\\
B_0^{(4)} &=& B_0(t_{12},m_b^2,m_b^2),\;\;\;\;
B_0^{(5)} = B_0(s_{23},m_b^2,m_W^2),\nn\\
C_\beta^{(1)} &=& C_\beta(0,t,0,m_b^2,m_W^2,m_b^2),\;\;
C_\beta^{(2)} = C_\beta(0,t,m_t^2,m_b^2,m_b^2,m_W^2),\nn\\
C_\beta^{(3)} &=& C_\beta(0,s_{23},0,m_b^2,m_W^2,m_b^2),\;\;
C_\beta^{(4)} = C_\beta(0,s_{23},m_t^2,m_b^2,m_b^2,m_W^2),\nn\\
C_\beta^{(5)} &=& C_\beta(0,t_{12},m_t^2,m_W^2,m_b^2,m_b^2),\;\;
C_\beta^{(6)} = C_\beta(0,0,t_{12},m_W^2,m_b^2,m_b^2),\nn\\
C_\beta^{(7)} &=&
C_\beta(t_{12},0,0,m_W^2,m_b^2,m_b^2),\;\beta=0,1,12,2,22,\nn\\
D_\lambda^{(1)} &=&
D_\lambda(0,0,0,m_t^2,s_{23},t_{12},m_W^2,m_b^2,m_b^2,m_b^2),\nn\\
D_\lambda^{(2)} &=&
D_\lambda(0,t_{12},0,t,m_t^2,0,m_W^2,m_b^2,m_b^2,m_b^2),\nn\\
\lambda &=& 0,1,11,12,13,2,22,23,3,33,003,112,113,122,123,133,222,
    223,233,333.\nn
\end{eqnarray}
The coefficient functions $C_{i,ij}$ and $D_{i,ij,ijk}$ are symmetric
functions and can further be decomposed into the scalar functions $A_0,
B_0, C_0,$ and $D_0$. See \cite{Denner}
for details.

\begin{thebibliography}{99}

\bibitem{Eilam:1990zc}
     G.~Eilam, J.~L.~Hewett and A.~Soni,
     Phys.\ Rev.\ D {\bf 44}, 1473 (1991)
     [Erratum-ibid.\ D {\bf 59}, 039901 (1999)].

\bibitem{Mele:1998ag}
    B.~Mele, S.~Petrarca and A.~Soddu,
    Phys.\ Lett.\ B {\bf 435}, 401 (1998)
    [arXiv:hep-ph/9805498].

\bibitem{Beneke:2000hk}
    M.~Beneke {\it et al.},
    arXiv:hep-ph/0003033. In this review one finds
that the reach of the LHC
is ${\rm Br}(t\to cg) \lsim 2.1\times 10^{-5}$
for integrated luminosity of $100\rm{fb}^{-1}$.
This limit would be impossible to achieve directly due
to a huge QCD background, see however \cite{ATLAS}, but rather through its
effect on the production of a single top quark via the parton
sub-process $gc\to t$.

\bibitem{ATLAS} J. Carvalho, N. Castro, A. Onofre
and F. Velosco (ATLAS Collaboration),
``Study of ATLAS sensitivity to FCNC top decays'', ATLAS internal note,
ATL-PHYS-PUB-2005-009, May 2005. Here the following limit is projected:
${\rm Br}(t\to cg) \lsim 4.3\times 10^{-4}$ for integrated luminosity of
$100{\rm fb}^{-1}$.

\bibitem{Cobal:2004zt}
    M.~Cobal,
    AIP Conf.\ Proc.\  {\bf 753}, 234 (2005)
    [arXiv:hep-ex/0412053].

\bibitem{Frank:2005vd}
     M.~Frank and I.~Turan,
     Phys.\ Rev.\ D {\bf 72}, 035008 (2005)
     [arXiv:hep-ph/0506197].

\bibitem{Arhrib:2005nx}
    A.~Arhrib,
    Phys.\ Rev.\ D {\bf 72}, 075016 (2005)
    [arXiv:hep-ph/0510107].

\bibitem{Jenkins:1996zd}
    E.~Jenkins,
    Phys.\ Rev.\ D {\bf 56}, 458 (1997)
    [arXiv:hep-ph/9612211].

\bibitem{Altarelli:2000nt}
    G.~Altarelli, L.~Conti and V.~Lubicz,
    Phys.\ Lett.\ B {\bf 502}, 125 (2001)
    [arXiv:hep-ph/0010090].

\bibitem{Bar-Shalom:2005cf}
    S.~Bar-Shalom, G.~Eilam, M.~Frank and I.~Turan,
c W W, c Z
    Phys.\ Rev.\ D {\bf 72}, 055018 (2005)
    [arXiv:hep-ph/0506167].

\bibitem{Bar-Shalom:1997sj}
    S.~Bar-Shalom, G.~Eilam, A.~Soni and J.~Wudka,
for t $\to$ c
    Phys.\ Rev.\ D {\bf 57}, 2957 (1998)
    [arXiv:hep-ph/9708358].

\bibitem{Greub:1996wn}
    C.~Greub, T.~Hurth, M.~Misiak and D.~Wyler,
    Phys.\ Lett.\ B {\bf 382}, 415 (1996)
    [arXiv:hep-ph/9603417].

\bibitem{Greub:2000sy}
     C.~Greub and P.~Liniger,
     Phys.\ Rev.\ D {\bf 63}, 054025 (2001)
     [arXiv:hep-ph/0009144].

\bibitem{Hou:1988wt}
     W.~S.~Hou,
     Nucl.\ Phys.\ B {\bf 308}, 561 (1988).

\bibitem{Hou:1990js}
     W.~S.~Hou and R.~G.~Stuart,
     Phys.\ Lett.\ B {\bf 242}, 467 (1990).

\bibitem{Simma:1990nr}
     H.~Simma and D.~Wyler,
     Nucl.\ Phys.\ B {\bf 344}, 283 (1990).

\bibitem{Liu:1989pc}
     J.~Liu and Y.~P.~Yao,
     Decay B
     Phys.\ Rev.\ D {\bf 41}, 2147 (1990).

\bibitem{Eilam:1989zm}
    G.~Eilam, B.~Haeri and A.~Soni,
    Phys.\ Rev.\ D {\bf 41}, 875 (1990).

\bibitem{Arhrib:2000ct}
    A.~Arhrib and W.~S.~Hou,
    Phys.\ Rev.\ D {\bf 64}, 073016 (2001)
    [arXiv:hep-ph/0012027].

\bibitem{Cordero-Cid:2004hk}
    A.~Cordero-Cid, J.~M.~Hernandez, G.~Tavares-Velasco and J.~J.~Toscano,
    arXiv:hep-ph/0411188.

\bibitem{Deshpande:1991pn}
   N.~G.~Deshpande, B.~Margolis and H.~D.~Trottier,
   Phys.\ Rev.\ D {\bf 45}, 178 (1992).

\bibitem{cutlers}
     R.~Cutler and D.~Sivers, Phys.\ Rev.\ D {\bf 17}, 196 (1978).

\bibitem{field}
     R.D. Field, ``Applications of Perturbative QCD,'' Perseus
     Books, 1989, Chapter 7.

\bibitem{'tHooft:1973mm}
     G.~'t Hooft,
     Nucl.\ Phys.\ B {\bf 61}, 455 (1973).

\bibitem{Hahn:2000jm}
    T.~Hahn and M.~Perez-Victoria,
    Comput.\ Phys.\ Commun.\  {\bf 118}, 153 (1999)
    [arXiv:hep-ph/9807565];
    T.~Hahn,
     Nucl.\ Phys.\ Proc.\ Suppl.\  {\bf 89}, 231 (2000)
     [arXiv:hep-ph/0005029];
     T.~Hahn, Comput.\ Phys.\ Commun.\ 140, 418 (2001);
     T.~ Hahn, C.~Schappacher, Comput.\ Phys.\ Commun.\ 143, 54 (2002);
     T.~Hahn,
     arXiv:hep-ph/0506201.

\bibitem{FF}
     G.~J.~van Oldenborgh, Comput.\ Phys.\ Commun.\ 66, 1 (1991);
     T. Hahn, Acta Phys.\ Polon.\ B 30, 3469 (1999).

\bibitem{Mertig:1990wm}
     R.~Mertig and J.~Kublbeck,
{\it Prepared for International Workshop on Software Engineering,
Artificial Intelligence and Expert Systems for High-energy and
Nuclear Physics,
Lyon, France, 19-24 Mar 1990};
     J.~Kublbeck, H.~Eck and R.~Mertig,
     Nucl.\ Phys.\ Proc.\ Suppl.\  {\bf 29A}, 204 (1992).

\bibitem{Kinoshita:1962ur}
     T.~Kinoshita,
     J.\ Math.\ Phys.\  {\bf 3}, 650 (1962).

\bibitem{Lee:1964is}
     T.~D.~Lee and M.~Nauenberg,
     Phys.\ Rev.\  {\bf 133}, B1549 (1964).

\bibitem{Sterman:1976jh}
     G.~Sterman,
     Phys.\ Rev.\ D {\bf 14}, 2123 (1976).

\bibitem{Passarino:1978jh}
     G.~Passarino and M.~J.~G.~Veltman,
     Nucl.\ Phys.\ B {\bf 160}, 151 (1979).

\bibitem{Fusaoka:1998vc}
    H.~Fusaoka and Y.~Koide,
    Phys.\ Rev.\ D {\bf 57}, 3986 (1998)
    [arXiv:hep-ph/9712201].

\bibitem{Aguilar-Saavedra:2002ns}
   J.~A.~Aguilar-Saavedra and B.~M.~Nobre,
   Phys.\ Lett.\ B {\bf 553}, 251 (2003)
   [arXiv:hep-ph/0210360].

\bibitem{Eilam:2006rb}
   G.~Eilam, M.~Frank and I.~Turan,
   arXiv:hep-ph/0601253.

\bibitem{Eilam:1991yv}
  G.~Eilam, J.~L.~Hewett and A.~Soni,
  Phys.\ Rev.\ Lett.\  {\bf 67}, 1979 (1991).

\bibitem{Denner}
     A.~Denner, Fortschr. Phys. {\bf 41} 307 (1993).

\end{thebibliography}

\end{document}